\documentclass[pre,aps,superscriptaddress,twocolumn]{revtex4-1}
\usepackage{graphicx,times}
\usepackage{bm}
\usepackage{amsmath}
\usepackage{amsfonts,subfigure,dsfont}
\usepackage{amssymb}
\usepackage{epsfig}
\usepackage{color}
\usepackage{amssymb,bbm,amsthm}

\begin{document}

\newcommand{\abs}[1]{\left\vert#1\right\vert}
\newcommand{\set}[1]{\left\{#1\right\}}
\newcommand{\bra}[1]{\left\langle#1\right\vert}
\newcommand{\ket}[1]{\left\vert#1\right\rangle}
\newcommand\braket[2]{\left.\left\langle#1\right|#2\right\rangle}
\newcommand{\avg}[1]{\left< #1 \right>}
\newcommand{\tr}[1]{\text{Tr}\left\{#1\right\}}
\def\I {{\rm 1} \hspace{-1.1mm} {\rm I} \hspace{0.5mm}}
\def\Z {{\mathds Z}}
\newcommand{\rosso}[1]{\color[rgb]{0.6,0,0} #1}
\newcommand{\GF}[1]{\textcolor{cyan}{#1}}

\title{Quantum Coherence and Ergotropy}
\author{G. Francica}
\affiliation{CNR-SPIN, I-84084 Fisciano (Salerno), Italy}
\author{F. C. Binder}
\affiliation{Institute for Quantum Optics and Quantum Information -- IQOQI Vienna, Austrian Academy of Sciences, Boltzmanngasse 3, 1090 Vienna, Austria}
\author{G. Guarnieri}
\affiliation{School of Physics, Trinity College Dublin, Dublin 2, Ireland.
}\author{M. T. Mitchison}
\affiliation{School of Physics, Trinity College Dublin, Dublin 2, Ireland.
}\author{J. Goold}
\affiliation{School of Physics, Trinity College Dublin, Dublin 2, Ireland.
}
\author{F. Plastina}
\affiliation{Dip. Fisica, Universit\`{a} della Calabria, 87036
Arcavacata di Rende (CS), Italy} \affiliation{INFN - Gruppo
Collegato di Cosenza}

\date{\today}

\begin{abstract}
Constraints on work extraction are fundamental to our operational understanding of the thermodynamics of both classical and quantum systems. In the quantum setting, finite-time control operations typically generate coherence in the instantaneous energy eigenbasis of the dynamical system. Thermodynamic cycles can, in principle, be designed to extract work from this non-equilibrium resource. Here, we isolate and study the quantum coherent component to the work yield in such protocols. Specifically, we identify a coherent contribution to the ergotropy (the maximum amount of unitarily extractable work via cyclical variation of Hamiltonian parameters). We show this by dividing the optimal transformation into an incoherent operation and a coherence extraction cycle. We obtain bounds for both the coherent and incoherent parts of the extractable work and discuss their saturation in specific settings. Our results are illustrated with several examples, including finite-dimensional systems and bosonic Gaussian states that describe recent experiments on quantum heat engines with a quantized load.
\end{abstract}

\bigskip
\pacs{ }

\maketitle

\section{Introduction}

The Thomson~\cite{Kelvin} formulation of the second law is a constraint on the ability of an external agent to extract work from a system. More precisely, it states that no work can be extracted from a closed equilibrium system during a cyclic variation of a parameter by an external source~\cite{thomson,uffink}. This formulation was influential in mathematical physics, leading to a definition of the condition of thermal equilibrium for quantum states through the notion of passivity and complete passivity~\cite{pusz,Lenard}. A state $\hat\rho$ is said to be passive with respect to a Hamiltonian when no work can be extracted from it by means of a cyclical variation of a Hamiltonian parameter, while it can be shown that a Gibbs state is the unique completely passive state such that $\hat\rho^{\otimes N}$ remains passive for all $N$. In other words, passivity allows us derive Thomson's formulation of the second law as a constraint on unitary work extraction from quantum systems~\cite{mathsecondlaw}. If a state is non-passive with respect to a Hamiltonian, work can be extracted and, upon maximization over the space of cyclical unitaries, the optimal yield is known as the {\it ergotropy}~\cite{ergotropy,adiabaticAvailability}. The ergotropy has been established as an important quantity in the emerging field of quantum thermodynamics~\cite{kosl,john,janet,Mitchison2019,Binder2018} and has recently been measured in two experiments which explore work deposition to external loads coupled to microscopic engines~\cite{Lindenfels2019, Horne2020}. In the limit of many copies, the ergotropy converges to the conventional non-equilibrium part of the free energy~\cite{Niedenzu2019} and it has also been incorporated into an open system thermodynamic description of finite quantum systems, recovering first and second laws~\cite{Binder2015}.

A central theme in the field of quantum thermodynamics over the last decade has been the identification of uniquely quantum signatures in thermodynamic settings. This includes the identification of quantum effects in thermal machines~\cite{scully,rahav,Brunner2014,Mitchison2015,delcampo,manycycles,Brandner2017,Klaers2017,Kilgour2018,holubec,klatzow,CampisiPRL,dann,jukka1,jukka2,peterson}, in work extraction protocols~\cite{funo,acin,skrzypczyk,pera,coh2w,alaff1,alaff2,lalo,morris,vitagliano,monsel,glg,daemon1,daemon2,Niedenzu2019}, in fluctuations of work~\cite{work0,work1,work2,w1,w2,w3,fullyquantum}, and in work deposition processes~\cite{alicki, binder, batteries, charging, andolina, bera, alioscia}, to name but a few examples. Arguably the most fundamental of all non-classical features is quantum coherence, yet precise mathematical techniques for its quantification have only recently been formulated in quantum information theory~\cite{plenio,cohrev}. From the perspective of quantum thermodynamics, many studies have aimed at highlighting the non-trivial role that coherence may play~\cite{Binder2018,coherenceaa,losta1,losta2,Uzdin2015,kamma,kallush,Archak2020NPJQI,guarnieriPLA,latune,cakmak2020ergotropy}. Coherence is a basis-dependent quantity that can be expressed in terms of the relative entropy between the state of the system at hand and its dephased counterpart in the relevant basis~\cite{plenio}. This provides a connection to the finite-time thermodynamics of quantum systems, where the relative entropy is ubiquitous in the assessment of irreversible entropy production of closed~\cite{donald,deffner,inner} and open systems~\cite{spohn1978,spohnlebowitz1978,esposito2010,deffner2,giacomo,camati,nazarov,espositon}. This connection was recently exploited in order to isolate a coherent contribution to the entropy production in quantum dynamics \cite{coherence19,santos19,Riechers2020,varizi}. Here, the relevant coherence is defined relative to the energy eigenbasis, which plays a distinguished role in thermodynamics.

In this work, we focus on the role of such coherence in ergotropic work extraction. We believe the simplicity of our approach together with its operational significance will be of particular interest to those interested in isolating non-classical signatures in quantum thermodynamics. We begin by introducing the basic notions of coherence and ergotropy in the following section. In Section~\ref{coherentErgSec}, we identify coherent and incoherent contributions to the ergotropy, while bounds for the coherent ergotropy are derived in Sec. \ref{boundsec}. We then provide examples to illustrate our results in Sec.~\ref{exasec} and, finally, summarise in Sec.~\ref{finasec}.

\section{Preliminaries}
\label{prelimSec}
Given a quantum system in an initial state $\hat{\rho}$, and a Hamiltonian $\hat H=\sum_k \varepsilon_k \ket{\varepsilon_k}\bra{\varepsilon_k}$, we are interested in the amount of coherence in the energy eigenbasis. In what follows, we will quantify the coherence with the \emph{relative entropy of coherence} $C(\hat{\rho})$~\cite{plenio,cohrev}. This is motivated from the description of coherence as a quantum resource theory~\cite{cohrev,Chitambar2019}. 

A quantum resource arises when there is a naturally restricted set of operations $\mathcal O$ which are significantly easier to implement than operations outside $\mathcal O$ -- e.g. local operations and classical communication (LOCC) in entanglement theory~\cite{Horodecki}. If these \emph{free operations} $\mathcal O$ only allow some \emph{free states} $\mathcal{F}$ to be created `for free', all other states become a resource whose creation requires the (costly) implementation of operations outside $\mathcal O$. We may quantify the resourcefulness of a non-free state by means of a function $\mu$ that maps states to non-negative reals. We call $\mu$ a \emph{resource monotone} if (i) its value cannot increase under application of any free operation $\Omega\in\mathcal O$ to any state $\hat\rho$: $\mu(\hat \rho)\geq\mu(\Omega(\hat\rho))$; and if (ii) $\mu(\hat \varphi)=0$ for all $\hat \varphi\in\mathcal F$. One way of constructing a monotone $\mu$ is to minimize a (contractive) distance function $d$ on the space of quantum states with respect to $\mathcal F$: $\mu_d(\hat \rho):=\min_{\hat \varphi\in\mathcal F}d(\hat \rho,\hat\varphi)$. The usefulness of such a distance-based $\mu_d$ then depends not least on its ease of computation -- i.e., if it can be expressed as a closed-form function.

Returning to coherence, various viable classes of free operations have been considered for which the free states $\mathcal F$ are the set of incoherent states $I_H$, i.e., density matrices $\hat \delta$ that are diagonal in the energy eigenbasis~\cite{cohrev}. For all of these classes, valid coherence monontones may be obtained based on suitable distance measures such as Tsallis relative $\alpha$-entropies $D_\alpha$ for which succinct expressions have been found~\cite{Rastegin2016}: $C_\alpha:=\min_{\hat \delta\in I_H}D_\alpha(\hat\rho||\hat\delta)$, where the normalized state $\hat\delta_{\rho,\alpha}\propto\sum\bra{\varepsilon_j}\hat\rho^\alpha\ket{\varepsilon_j}^{1/\alpha}\ket{\varepsilon_j}\bra{\varepsilon_j}$ obtains the minimum. We here focus on the limit $\alpha\to 1$ as, in this case, the minimal state $\hat\delta_\rho\equiv\hat\delta_{\rho,\alpha}=\Delta(\hat\rho)$ is directly connected to the original state $\hat \rho$ by a physical operation -- dephasing with $\Delta$. In this limit, $D_\alpha$ becomes the standard quantum relative entropy $D(\hat{\rho}||\hat{\delta})= \tr{\hat{\rho} (\log \hat{\rho} - \log \hat{\delta})}$ and $C_\alpha$ becomes the entropy of coherence $C(\hat \rho)=S(\hat{\delta}_{\rho}) - S(\hat{\rho})$, with the Von Neumann entropy $S(\hat{\sigma}) = - \tr{\hat{\sigma} \, \log \hat{\sigma}}$~\cite{plenio}.

Following the seminal paper~\cite{ergotropy}, we are now interested in extracting work from the quantum system at hand by using a cyclic unitary transformation $\hat U \in {\cal U}_c$, where ${\cal U}_c$ denotes the set of unitary transformations generated in a given interval $(0,\tau)$ by a time dependent Hamiltonian $\hat{H}(t)$ such that $\hat H(0) = \hat H(\tau) = \hat H$. In this context, one typically assumes complete control over the system~\cite{kallush}: that is, the possibility of generating any unitary evolution through suitable control fields applied to the system, which are switched off at the end of the transformation. Under the action of the unitary $\hat{U}$, the state transforms as $\hat{\rho} \rightarrow \hat U \hat{\rho} \hat U^{\dag}$, and the average work extracted from the system is $W (\hat{\rho}, \hat U) = \tr{\hat H (\hat{\rho} - \hat U \hat{\rho} \hat U^{\dag}) }$. The maximum of $W$ over the set ${\cal U}_c$ is called ergotropy, $\cal E$. After ordering the labels of eigenstates of $\hat H$ and of $\hat{\rho}$ such that $\hat H = \sum_{k=1}^d \varepsilon_k \ket{\varepsilon_k}\bra{\varepsilon_k}$, with $\varepsilon_k < \varepsilon_{k+1}$, and  $\hat{\rho} = \sum_{k=1}^d r_k \ket{r_k}\bra{r_k}$, with $r_k \geq r_{k+1}$, we define the optimal {\it ergotropic} transformation $\hat E_{\rho}$ as the one that maps $\hat{\rho}$ into the passive state $\hat P_{\rho} = \hat E_{\rho} \hat{\rho} \hat{E}_{\rho}^{\dag} = \sum_k r_k \, \ket{\varepsilon_k} \bra{\varepsilon_k}$. We notice that the optimal unitary $\hat E$ depends on the state $\hat{\rho}$, and that the ergotropy is then given by
\begin{eqnarray} {\cal E}(\hat{\rho}) &= & \mbox{max}_{\hat U \in {\cal U}_c} W(\hat{\rho}, \hat U) \equiv W(\hat{\rho}, \hat E_{\rho}) =  \tr{ \hat H (\hat{\rho} - \hat P_{\rho}) }\nonumber \\ & \equiv & \sum_{k} \varepsilon_k (\rho_{kk} - r_k) \, ,\end{eqnarray}
where $\rho_{kk}$ (the population of $\hat{\rho}$ in the $k$-th energy eigenstate) can be expressed as $\rho_{kk} = \sum_{k'} r_{k'} |\langle r_{k'}\ket{\varepsilon_k}|^2$. Our main aim is to demonstrate a precise connection between $\cal E$ and the amount of coherence in the initial state $C(\hat{\rho})$~\cite{degeneracy}. In the following section, we show how to split the ergotropy into two contributions, one of which directly connected to the presence of energetic coherence in the state $\hat{\rho}$.

\section{Coherent and incoherent contributions to ergotropy}
\label{coherentErgSec}

We start by introducing the incoherent part of the ergotropy, ${\cal E}_i$, which can be defined in two equivalent ways. One can think of ${\cal E}_i$ as the maximum work extractable from $\hat{\rho}$ without altering its coherence. To formalize this idea, we can imagine breaking the transformation $\hat E_{\rho}$ into an incoherent operation followed by a second, coherence-consuming, cyclic unitary. To this end, we define the subset ${\cal U}_c^{(i)} \subset {\cal U}_c$ of incoherent, cyclic, unitary transformations, such that any $\hat V \in {\cal U}_c^{(i)}$ is coherence-preserving: $C(\hat{\rho}) = C(\hat V\hat{\rho} \hat V^{\dag})$. Such $\hat V$ is in fact a member of the class of \emph{strictly incoherent operations} (SIOs) which admit a very operational structure~\cite{Yadin2016,Peng16,cohrev}; $\hat V$ amounts to a  reshuffling of the energy basis, up to irrelevant phase factors, of the form $\hat V = \sum_k e^{- i \varphi_k} \ket{\varepsilon_{k}} \bra{\varepsilon_{\pi_k}} \equiv \hat V_{\pi}$, where $\pi_k$ is the $k$-th element in the result of the permutation $\pi$ of the indices~\cite{fixperm}. The incoherent contribution to ergotropy is then defined as
\begin{equation} {\cal E}_i = \mbox{max}_{\hat V \in {\cal U}_c^{(i)}} W(\hat{\rho}, \hat V) \equiv \mbox{max}_{\pi} W(\hat{\rho}, \hat V_{\pi}) \, .\end{equation}
The optimal permutation, $\tilde{\pi}$, realizing the maximum in the equation above, is the one that rearranges the populations $\{\rho_{kk}\}_{k=1,\ldots d}$ in descending order: $\rho_{\tilde{\pi}_j \tilde{\pi}_j} \geq \rho_{\tilde{\pi}_{j+1} \tilde{\pi}_{j+1}}\, , \forall j$. Letting $\hat{\sigma}_{\rho} = \hat V_{\tilde{\pi}} \hat{\rho} \hat V_{\tilde{\pi}}^{\dag} = \sum_k \sum_{k'} \rho_{\tilde{\pi}_k,\tilde{\pi}_{k'} } \ket{\varepsilon_k} \bra{\varepsilon_{k'}}$, the incoherent contribution to ergotropy is
\begin{equation}
{\cal E}_i( \hat{\rho}) = \tr{\hat H (\hat{\rho} -\hat{\sigma}_{\rho} )} = \sum_k \varepsilon_k (\rho_{kk} - \rho_{\tilde{\pi}_k \tilde{\pi}_k} ). \label{tre} \end{equation}
The state $\hat{\sigma}_{\rho}$ possesses the same coherence as $\hat{\rho}$, but less average energy. Therefore, ${\cal E}_i$ is the maximum amount of work that can be extracted from $\hat{\rho}$ without changing its coherence, and, among the states having the same amount of coherence as $\hat{\rho}$, $\hat{\sigma}_{\rho}$ is singled out as the one that possesses the least possible average energy~\cite{nota}. In particular, we notice that, when trying to extract work from the state $\hat{\sigma}_{\rho}$ through the optimal cyclic unitary $\hat E_{\sigma}$, one arrives at the very same passive state that is obtained from $\hat{\rho}$. In our notation, $\hat P_{\sigma} = \hat P_{\rho}$. This is because $\hat{\sigma}_{\rho}$ has the same eigenvalues as $\hat{\rho}$.

An alternative (but equivalent) route to the identification of the incoherent contribution to ergotropy is provided by defining ${\cal E}_i$ as the maximum amount of work extractable from $\hat{\rho}$ after having erased all of its coherences via the dephasing map $\Delta$. This amounts to defining ${\cal E}_i$ as the full ergotropy of the dephased state, ${\cal E}_i = {\cal E}(\hat{\delta}_{\rho})$, where $\hat{\delta}_{\rho} = \Delta[\hat{\rho}]$ has the same energy populations as $\hat{\rho}$ (and, thus, the same average energy) but zero coherence. The ergotropy of $\hat{\delta}_{\rho}$ can be written by first defining the passive state $\hat P_{\delta}$ obtained from $\hat{\delta}_{\rho}$ after re-arranging the populations in decreasing order, and then letting
\begin{equation}
{\cal E}_i(\hat{\rho}) \equiv {\cal E}(\hat{\delta}_{\rho}) = \tr{\hat H ( \hat{\delta}_{\rho} - \hat P_{\delta})} \, . \label{cinque}
\end{equation}
This definition is fully equivalent to the one given in Eq.~\eqref{tre}~\cite{equivalence}. Indeed, $\hat{\delta}_{\rho}$ has the same populations as $\hat{\rho}$ in the energy basis; consequently, the optimal reshuffling unitary that maps $\hat{\delta}_{\rho}$ into $\hat P_{\delta}$ is given by the very same $\hat V_{\tilde{\pi}}$ introduced above. This implies that $\hat P_{\delta}$ has the same populations as $\hat{\sigma}_{\rho}$ (in the same order!), but no coherence. As a result of these considerations, one immediately realizes that $\hat P_{\delta}$ can be obtained by applying the dephasing map to $\hat{\sigma}_{\rho}$, and that the two states share the same average energy:
$$\hat P_{\delta} \equiv \Delta[\hat{\sigma}_{\rho}] \quad \Rightarrow \quad \tr{\hat H \hat{\sigma}_{\rho}}\equiv \tr{\hat H \, \hat P_{\delta}}\, .$$
Having defined the incoherent part of ${\cal E}(\hat{\rho})$, the coherent contribution to ergotropy is simply given by the difference
\begin{equation}
{\cal E}_c = {\cal E} - {\cal E}_i = \tr{\hat H (\hat{\sigma}_{\rho} - \hat P_{\rho})} \equiv \sum_k \varepsilon_k (\rho_{\tilde{\pi}_k\tilde{\pi}_k} - r_k) \, .\end{equation}
This is a non-negative quantity as, in general, $\hat{\sigma}_{\rho}$ is an active state. Notice further, that it coincides with the full ergotropy of $\hat{\sigma}_{\rho}$.

The coherent ergotropy $\mathcal{E}_c$ can be understood as that part of extractable work which cannot be obtained by means of incoherent operations applied to state $\hat{\rho}$, and it is due to the presence of coherence in the initial state. Despite this, ${\cal E}_c$ is not a coherence monotone, as the inequality ${\cal E}_c( \hat V \hat{\rho} \hat V^{\dag}) \leq {\cal E}_c(\hat \rho) $ is not satisfied for every incoherent operation $\hat V$ (see Appendix for an illustrative example). Nevertheless, both the state $\hat{\sigma}_{\rho}$ and the coherent part of the ergotropy, ${\cal E}_c$, are uniquely defined once the state $\hat{\rho}$ and the Hamiltonian $\hat{H}$ are given, and they result entirely from the initial coherence, implying that $\hat{\sigma}_{\rho}$ is not passive.

\begin{figure}
        \begin{center}
     \includegraphics[width=\columnwidth]{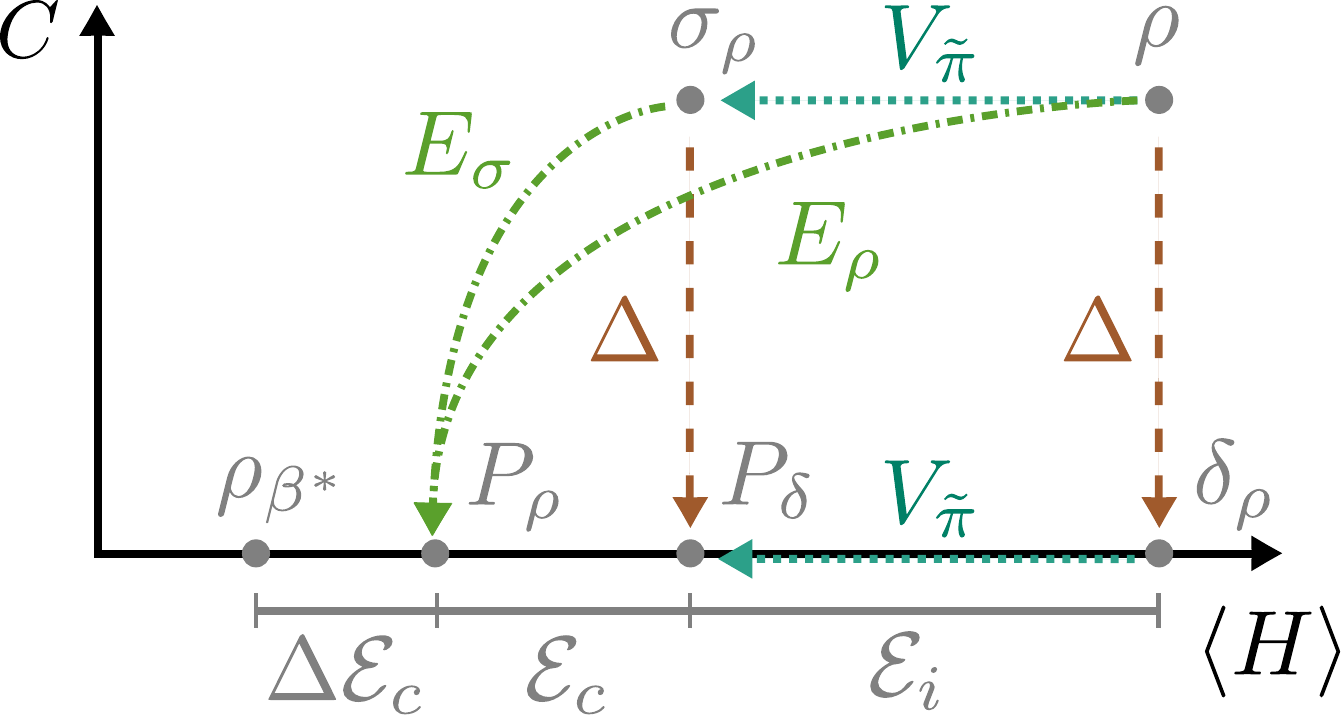}
\caption{Position of the various states (see main text) in a coherence-versus-average-energy diagram. Grey dots represent quantum states -- e.g. arising from the initial state $\hat{\rho}$ after the transformations $\hat E_\rho, \Delta, \hat V_{\tilde{\pi}}$ are performed. The arrows representing transformations are intended merely to point from the initial to the final state, without implying a precise path in the plane. For example, the transformation $\hat V_{\tilde{\pi}}$ is represented by a horizontal line because it connects states with the same amount of coherence; however, coherence may change {\it during} the transformation. The horizontal distance $\Delta {\cal E}_c$ between the thermal state $\hat{\rho}_{\beta^*}$ and $\hat P_{\rho}$ is the \emph{bound ergotropy} (see Sec.~\ref{boundsec}). It may be zero, depending on the system at hand (i.e., iff the eigenvalues of $\rho$ and $\rho_{\beta^*}$ are related by a permutation).}
        \label{schemaCH}
        \end{center}
\end{figure}

Fig.~\ref{schemaCH} summarizes these considerations and relationships. It shows the various states and operations defined up to now in the coherence-versus-average-energy plane.

\section{Bounds for coherent ergotropy}
\label{boundsec}
Given the form of the coherent ergotropy, we can provide upper and lower bounds to its value and show their tightness. Indeed, by introducing the Gibbs state $\hat{\rho}_{\beta} = e^{-\beta \hat H}/Z$ with inverse temperature $\beta$, we can exploit the identity $D(\hat{\sigma}||\hat{\rho}_{\beta}) = \beta \tr{\hat H ( \hat{\sigma} - \hat{\rho}_{\beta})} - S(\hat{\sigma}) + S(\hat{\rho}_{\beta})$, valid for any state $\hat{\sigma}$, in order to obtain the following chain of relations:
\begin{eqnarray*}
\beta {\cal E}_c &= &\beta ({\cal E}- {\cal E}_i) = \beta \tr{\hat H \left (\hat{P}_{\delta} - \hat P_{\rho} \right )} = \\
&=& \beta \tr{\hat H \left (\hat{P}_{\delta} - \hat{\rho}_{\beta} \right )} - \beta \tr{\hat H \left (\hat{P}_{\rho} - \hat{\rho}_{\beta} \right )} = \\
&=&\left [ D(\hat P_{\delta}|| \hat{\rho}_{\beta}) + S(\hat P_{\delta}) - S(\hat{\rho}_{\beta}) \right ]  \\
 && -\left [ D(\hat P_{\rho}|| \hat{\rho}_{\beta}) + S(\hat P_{\rho}) - S(\hat{\rho}_{\beta}) \right ] \end{eqnarray*}
After taking into account that $ S(\hat P_{\rho})= S(\hat{\rho})$, and that $ S(\hat P_{\delta})= S(\hat{\delta}_{\rho})$ (due to the fact that they are connected by unitary transformations), and, finally, using $C(\hat{\rho}) = S(\hat{\delta}_{\rho})-S(\hat{\rho})$, we obtain
\begin{equation}\label{coherentergotropyid}
\beta {\cal E}_c = C(\hat{\rho}) + D(\hat P_{\delta}|| \hat{\rho}_{\beta}) - D(\hat P_{\rho}|| \hat{\rho}_{\beta}) \, ,
\end{equation}
which is valid for every finite $\beta$.

From this relation, using the fact the $D \geq 0$, one easily obtains bounds for ${\cal E}_c(\hat{\rho})$:
\begin{equation}
C(\hat{\rho}) - D(\hat P_{\rho}|| \hat{\rho}_{\beta}) \leq \beta {\cal E}_c (\hat{\rho}) \leq C(\hat{\rho}) + D(\hat P_{\delta}|| \hat{\rho}_{\beta}) \, . \label{bounds}
\end{equation}
One can saturate the upper bound if $\hat P_{\rho} = \hat{\rho}_{\beta}$. This requires that the ergotropic transformation $\hat E_{\rho}$ takes $\hat{\rho}$ to the thermal state $\hat{\rho}_{\beta}$. Due to unitarity of this transformation, a necessary condition on $\beta$ is that $S(\hat{\rho}) = S(\hat{\rho}_{\beta^*})$. We label the specific value of $\beta$ for which this entropic equality holds $\beta^*$, and note that it exists for any $\hat\rho$. Moreover, for a single qubit, as well as for the important class of bosonic or fermionic states of Gaussian form, the condition $\beta=\beta^*$ is not just necessary, but also sufficient for the saturation of the upper bound in Eq.~(\ref{bounds}) (see examples in Sec.~\ref{exasec}).

More generally, however, the choice $\beta=\beta^*$ does not imply saturation of the bound. That is, the difference
\begin{align}
\begin{split}
\Delta \mathcal{E}_c&:= \frac{1}{\beta^*}\left[ C(\hat{\rho}) + D(\hat P_{\delta}|| \hat{\rho}_{\beta^*})\right]- {\cal E}_c (\hat{\rho})\\
&= \frac{1}{\beta^*} D(\hat P_\rho||\hat \rho_{\beta^*})\geq 0 
\end{split}
\label{DeltaE}
\end{align}
does not generally vanish. In fact, by expressing it as $\Delta \mathcal{E}_c  = \tr{\hat H(\hat P_\rho-\hat{\rho}_{\beta^*})}$ we note that it equates to what is called the \emph{bound ergotropy} $\mathcal{E}_{b}$~\cite{Niedenzu2019} -- i.e., the amount of additional ergotropy that a global unitary transformation could retrieve from $\hat{\rho}^{\otimes n}$, per system, in the limit $n\to\infty$ (in addition to the single-system ergotropy $\mathcal{E}$).

The saturation of the upper bound of Eq.~(\ref{bounds}) is, furthermore, equivalent to the results of Ref.~\cite{coherence19} where the irreversible work $W_{irr}$ {\it performed} on a quantum system was analyzed for a unitary transformation taking an initial thermal state $\hat{\rho}_{\beta^*}$ to a final state $\hat{\rho}= \hat U \hat{\rho}_{\beta^*} \hat U^{\dag}$. It was shown that $\beta^*W_{irr}=C(\hat{\rho})+D(\hat{\delta}_{\rho}|| \hat{\rho}_{\beta^*})$. For a cyclic transformation, $W_{irr}$ coincides with the average work performed on the system, whose absolute value, in turn, coincides with the work extracted from it by the cyclic unitary $\hat U^{\dag}$, when it is prepared in the state $\hat{\rho}$. If we take $\hat U^{\dag} = \hat{E}_{\rho}$, then the result of Ref.~\cite{coherence19} is translated into our notation as
\begin{equation}\beta^* {\cal E}(\hat{\rho}) = C(\hat{\rho}) + D(\hat{\delta}_{\rho}|| \hat{\rho}_{\beta^*}), \; \mbox{if } \, \hat E_{\rho} \hat{\rho} \hat E_{\rho}^{\dag} = \hat{\rho}_{\beta^*} \, . \label{vecchioris}\end{equation}
But, with the same argument as given above, the incoherent ergotropy, Eq.~(\ref{cinque}), can be rewritten (for any $\beta$) as
\begin{equation}
\beta {\cal E}_i (\hat{\rho}) = D(\hat{\delta}_{\rho}|| \hat{\rho}_{\beta}) -  D(\hat P_{\delta}|| \hat{\rho}_{\beta}) \, .
\end{equation}
Taking $\beta=\beta^*$, and subtracting this relation from Eq.~(\ref{vecchioris}), we obtain the saturation of the upper bound of Eq.~(\ref{bounds}):
\begin{equation} \beta^* {\cal E}_c(\hat{\rho}) = C(\hat{\rho}) + D(\hat P_{\delta}|| \hat{\rho}_{\beta^*}), \; \mbox{if } \, \hat E_{\rho} \hat{\rho} \hat{E}_{\rho}^{\dag} = \hat{\rho}_{\beta^*} \, . \end{equation}

The lower bound in Eq.~(\ref{bounds}), on the other hand, is saturated iff $\hat P_{\delta}$ = $\hat{\rho}_{\beta}$ for some inverse temperature $\beta$. For $\mathcal{E}_c>0$, this requires that the populations of $\hat{\rho}$ in the energy basis (coinciding with those of $\hat{\delta}_{\rho}$) are indeed thermal, but in the wrong order, and that the state $\hat{\rho}$ contains some coherence in the energy basis. An example is provided by the following qutrit density matrix, written in the energy basis:
$$\hat{\rho}=\begin{pmatrix}
         g_1 & c & 0 \\
         c^* & g_3 & 0 \\
         0 & 0 & g_2
       \end{pmatrix}, \quad g_i =\frac{e^{-\beta \varepsilon_i}}{\sum_j e^{- \beta \varepsilon_j}}, \; |c| \leq \sqrt{g_1g_3} \, .$$
For such a state, the three populations $r_i$ are obtained by decreasingly ordering the set of numbers $\left\{\frac{g_1+g_3}{2} + \sqrt{\frac{(g_1-g_3)^2}{4} + |c|^2}; g_2; \frac{g_1+g_3}{2} - \sqrt{\frac{(g_1-g_3)^2}{4} + |c|^2}\right\}$, and the passive state $\hat P_{\rho}$ is obtained by taking the ordered set as energy level populations. On the other hand, $\hat P_{\delta} \equiv \hat{\rho}_{\beta}=\mbox{diag}\{g_1,g_2,g_3\}$; but this thermal state does not have the same entropy as $\hat{\rho}$ (and $\beta$ has nothing to do with $\beta^*$).
Using the definitions above, we obtain  ${\cal E} = \varepsilon_1 (g_1 - r_1) + \varepsilon_2 (g_3 - r_2) +\varepsilon_3 (g_2 - r_3)$, while ${\cal E}_i= (\varepsilon_3-\varepsilon_2) (g_2-g_3)$. The difference between these two quantities gives ${\cal E}_c$, which saturates the lower bound in Eq.~(\ref{bounds}) (i.e., for these states, $D(\hat P_\delta||\hat \rho_\beta)=0$).

Lastly, we can exploit Eq.~\eqref{coherentergotropyid} to investigate the convertibility of the states $\hat P_{\delta}$ and $\hat P_{\rho}$ under thermal operations, and endow this problem with an operational interpretation thanks to the definition of ergotropy.
Since both these states commute with the Hamiltonian and are passive, their convertibility may be addressed within the resource theory of athermality~\cite{Janzing2000,Brandao2013,Brandao2015}.
In particular, if a thermal operation~\cite{Brandao2013,Ng2019} exists that takes $\hat P_{\delta}$ to $\hat P_{\rho}$ ($\hat P_{\rho}$ to $\hat P_{\delta}$), it follows that $D(\hat P_{\delta}|| \hat{\rho}_{\beta}) - D(\hat P_{\rho}|| \hat{\rho}_{\beta}) \equiv \beta \mathcal E_c - C(\hat \rho) \geq 0$ ($\leq 0$, respectively). 

\section{Examples}
\label{exasec}

\subsection{Qudits}
In order to illustrate our results, we consider first the simple case of a qubit, having energy eigenvalues $\varepsilon_1=0$ and  $\varepsilon_2$. In this case, any initial state $\hat{\rho}$ is transformed by the ergotropic transformation $\hat E$ into a passive state with a thermal structure $\hat P_{\rho}\equiv \hat{\rho}_{\beta^*}$, for a suitably chosen inverse temperature $\beta^*$. Then, $\Delta\mathcal{E}_c$ vanishes and the upper bound in Eq.~(\ref{bounds}) is saturated. Moreover, in this case, the coherent part of ergotropy can be directly expressed in terms of the purity of the state, $p(\hat{\rho}) = \tr{\hat{\rho}^2}$ and of another coherence quantifier, the $l_1$ norm of coherence~\cite{cohrev}, defined as $C_{l_1}(\hat{\rho})= 2 \abs{\bra{\varepsilon_1} \hat{\rho}\ket{\varepsilon_2}}$. Indeed, some simple manipulations lead to

\begin{align}
    {\cal E}_c(\hat{\rho}) = \frac{\varepsilon_2}{2}\left( \sqrt{2 p(\hat{\rho})-1}-\sqrt{2 p(\hat{\rho})-1-C_{l_1}^2(\hat{\rho})}\right).
    \label{eq:Ec_qubit}
\end{align}
This is proved by noticing that ${\cal E}_c(\hat{\rho})=\varepsilon_2 (\rho_{22}-r_2)$, where the smallest eigenvalue of $\hat{\rho}$ is $r_2=\frac{1}{2}(1-\sqrt{2 p(\hat{\rho})-1})$, and where the smallest population of $\hat{\rho}$ is $\rho_{22}=\frac{1}{2}-\frac{1}{2}\sqrt{2p(\hat{\rho}) - 1 -C_{l_1}^2}$.

It follows from Eq.~\eqref{eq:Ec_qubit} that the ergotropy increases for any operation $\Omega$ with $p(\Omega(\hat{\rho}))<\frac{1}{2}+\frac{1}{2}\left(\frac{{\cal E}_c(\hat{\rho})}{\varepsilon_2}+\frac{1}{4}C_{l_1}^2(\Omega(\hat{\rho}))\frac{\varepsilon_2}{{\cal E}_c(\hat{\rho})}\right)^2$. In the Appendix we provide an example of an incoherent such operation -- generalized amplitude damping -- to prove that $E_c$ is not a coherence monotone.

For a given value of the purity $p$, the coherence takes its maximum value for mixed states $\hat{\rho}$ with equal populations, $\rho_{11}=\rho_{22}=1/2$, for which  $p=(1+C_{l_1}^2)/2$ and  ${\cal E}_c = C_{l_1}/2$. It follows that ${\cal E}_c(\rho)$ is maximized if the initial state is a maximally coherent pure state with $C_{l_1}=1$ and $p=1$.

This latter observation is, in fact, more general: for a $d$-level system, we get the maximum value of ${\cal E}_c(\hat{\rho})$ (with, correspondingly, a null incoherent contribution ${\cal E}_i$)  when $\hat{\rho}$ is a maximally coherent pure state, $\hat{\rho}=\ket{\psi}\bra{\psi}$, with $\ket{\psi}= \sum_i \ket{\varepsilon_i}/\sqrt{d}$. In such a case, indeed, any incoherent unitary $\hat{V}_{\pi}$ preserves the average energy. 

\begin{figure}
  \centering
  \includegraphics[width=0.8\columnwidth]{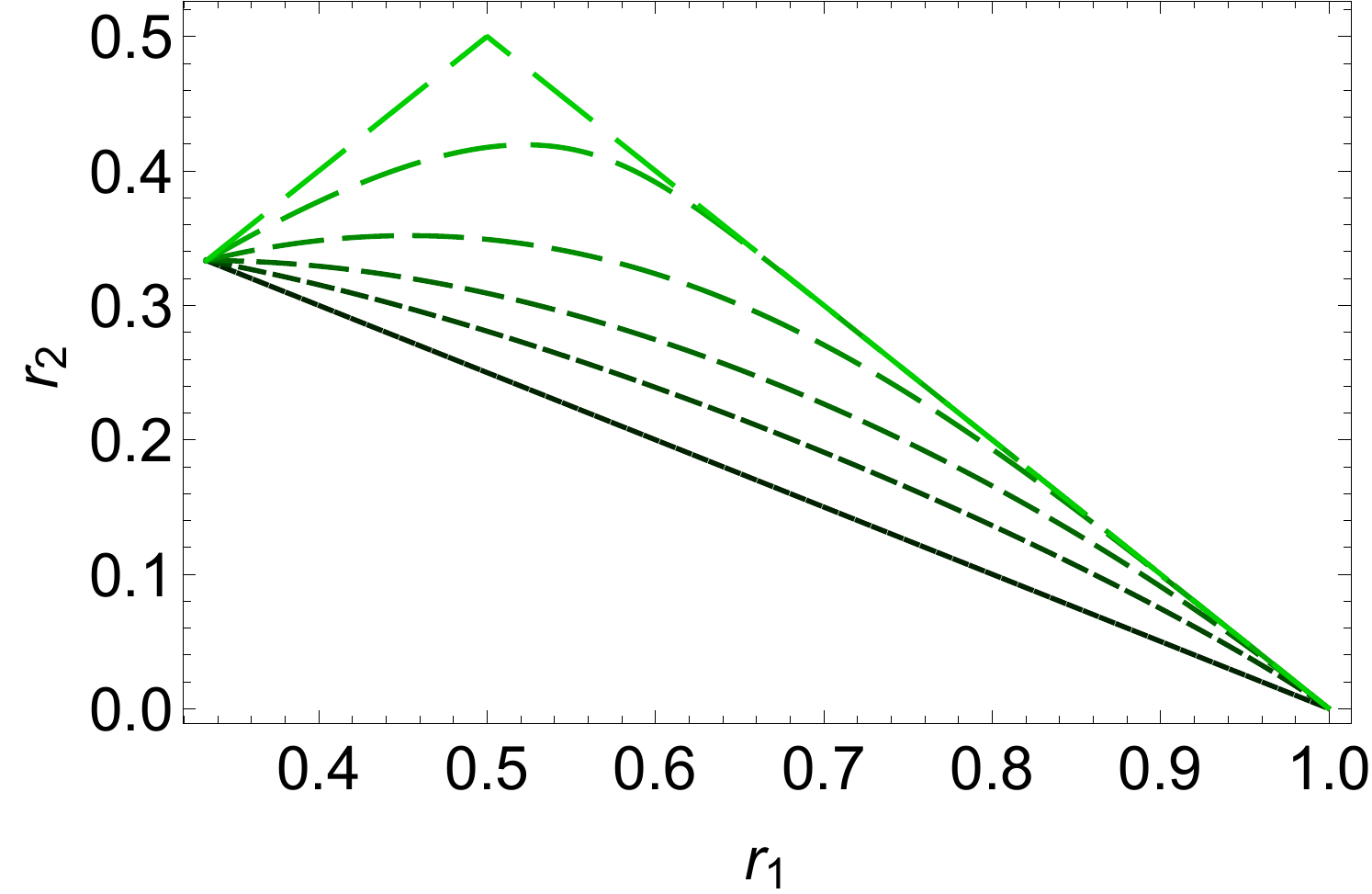}
  \caption{In a three-level system, we identify the class of states which allow to saturate the right inequality in Eq.~(\ref{bounds}) by looking at a pair of eigenvalues, $r_1$ and $r_2$, for which $\Delta {\cal E}_c=0$. The various lines refer to the cases in which the ratio of the second and third energy eigenvalues is given by $R=0,0.1,0.3,0.5,0.7,1$ (lighter dashed to darker solid lines).
  }\label{fig:diff}
\end{figure}

To discuss a less trivial case, where the upper bound in Eq.~(\ref{bounds}) is not always saturated, we now consider the behavior of the coherent part of ergotropy for a three-level system with energy eigenvalues $\varepsilon_1=0$, and $\varepsilon_2 =R \, \varepsilon_3$ (with $R \in (0,1)$). In particular, we ask under what conditions the bound is saturated (i.e., $\Delta\mathcal{E}_c=0$). Selecting $\beta=\beta^*$ as required for saturation, Eq.~\ref{DeltaE} implies that once the energy values are fixed, what really matters are just the first two eigenvalues of the density matrix, $r_1, r_2$ (which fix the third one as $r_3=1-r_1-r_2$). For our three-level system, the bound ergotropy can be written as $\Delta {\cal E}_c = \varepsilon_3  [r_2 (R-1)+ 1- r_1  - Z^{-1}(R e^{-\beta^* R \varepsilon_3} + e^{-\beta^* \varepsilon_3})  ]$, where $Z= 1 + e^{-\beta^* R \varepsilon_3} + e^{-\beta^* \varepsilon_3}$.
Looking for the values of $r_1$ and $r_2$ that give rise to a vanishing $\Delta {\cal E}_c$, we obtain the numerical results reported in Fig.~\ref{fig:diff}, where we can appreciate that only under very stringent conditions on the eigenvalues of $\hat{\rho}$ one obtains a saturation of the inequality. For fixed $R$, all suitable eigenvalue pairs are confined to a single curve within the total $(r_1,r_2)$-plane.

\subsection{Bosonic Gaussian states}

Beyond finite-dimensional systems, our results can also be directly applied to bosonic Gaussian states. These states arise naturally in the description of weakly interacting fermions or bosons and are, by definition, related to a thermal state by a unitary transformation. As a consequence, they saturate the upper bound in Eq.~\eqref{bounds}.

Let us focus for simplicity on a single bosonic mode, with Hamiltonian $\hat{H} = \hbar\omega \hat{a}^\dagger \hat{a}$, which is assumed to be in a Gaussian state of the form
\begin{equation}
    \label{dispTherm}
\hat{\rho} = \hat{D}(\alpha) \hat{\rho}_\beta \hat{D}^\dagger(\alpha),
\end{equation}
where $\hat{D}(\alpha) = e^{\alpha\hat{a}-\alpha^*\hat{a}^\dagger}$ is a unitary displacement operator. This could describe, for example, the mechanical motion of the trapped-ion heat engine reported in Ref.~\cite{Lindenfels2019}, whose vibrations act as a load or ``flywheel'' driven by a two-level working medium comprising the ion's electronic spin states. In that context, the displacement $\alpha$ arises from mechanical work performed by the engine on the load, while the thermal occupation $\bar{n}= (e^{\beta\hbar\omega}-1)^{-1}$ is associated with random energy transfer due to thermal fluctuations of the working medium. The total energy of such a state is then given by $E  = {\rm Tr}[\hat{H}\hat{\rho}] = \hbar\omega( |\alpha|^2 + \bar{n})$, while the total ergotropy is given simply by $\mathcal{E} =\hbar\omega |\alpha|^2$. We note that, since the dephasing operation $\Delta$ is non-Gaussian, it is difficult to obtain a simple closed-form expression for $\mathcal{E}_{c}$, but it can be readily computed numerically for small $|\alpha|$ and $\bar{n}$.

\begin{figure}
\centering
\includegraphics[width = 0.7\linewidth]{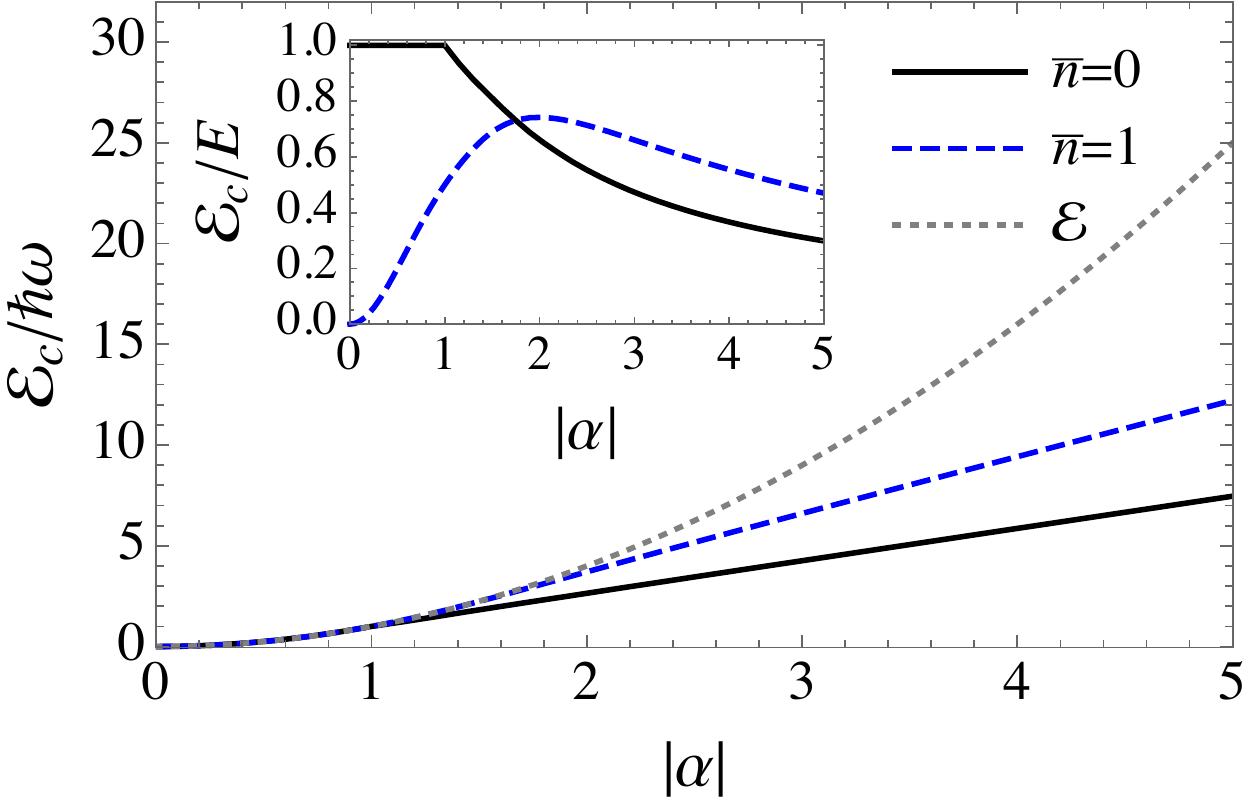}
\caption{Coherent ergotropy plotted in units of energy quanta (main) and as a fraction of the total energy (inset), for a displaced thermal state [Eq.~\eqref{dispTherm}] of a single bosonic mode as a function of the displacement, $\alpha$. The solid black line shows a pure coherent state, the dashed blue line shows a state with a thermal occupation of $\bar{n}=1$, while the dotted grey line shows the total ergotropy (equal for both states). \label{fig:disp_therm}}
\end{figure}

Fig.~\ref{fig:disp_therm} displays the coherent part of the ergotropy evaluated for two different examples: a pure coherent state with $\bar{n}=0$, and a displaced thermal state with $\bar{n}=1$. For $\alpha \ll 1$, the population distribution (i.e., the dephased state $\hat{\delta}_\rho$) is passive and therefore all the ergotropy is coherent, i.e., $\mathcal{E}_c = \mathcal{E}$. Conversely, for large $\alpha$, the coherent ergotropy is linear in the coherent displacement, $\mathcal{E}_c\propto |\alpha|$, while the total ergotropy is quadratic, $\mathcal{E}\propto |\alpha|^2$. Therefore, the energetics of Gaussian states with large displacement is dominated by the incoherent ergotropy, which is consistent with the quasi-classical nature of these states. The work content of such states derives primarily from the non-passivity of the population distribution. 

Interestingly, increasing $\bar{n}$ for fixed $\alpha$ actually increases the coherent ergotropy. This is because, for a fixed value of $\alpha$, thermal noise renders the population distribution more passive, thus decreasing $\mathcal{E}_i$ without changing the total ergotropy. This does not conflict with the obvious fact that, for fixed energy $E$, increasing $\bar{n}$ must imply that $|\alpha|$ is smaller and therefore both components of the ergotropy are reduced. 

\section{Summary and conclusions}
\label{finasec}

In summary, in this paper we have highlighted the role of quantum coherence in work extraction processes, by identifying a contribution to the ergotropy that precisely corresponds to initial coherence in the energy basis. This is obtained by breaking the optimal, ergotropic, unitary cycle into an initial incoherent unitary operation, followed by a second unitary cycle through which one extracts work by exhausting the coherence. We have analyzed this coherent ergotropic contribution by exploring its range of possible values, which we have identified in terms of two bounds which can be saturated in specific cases. In particular, we discovered that the tightness of the upper bound is intimately related to the concept of bound ergotropy -- a form of work potential that becomes available only when processing multiple identical copies of the system together. Finally, we have illustrated our results with the simplest non-trivial examples of a qubit and a qutrit, as well as a single-mode bosonic Gaussian state. The latter opens the possibility for future analysis of work extraction in continuous variable systems beyond unconstrained unitaries on single modes, considering, for instance, Gaussian operations, multiple modes, or both~\cite{Brown2016,Friis2018,Singh2019,Narasimhachar2019,Serafini2020}.

As quantum coherence is arguably the most primordial non-classical effect in nature, we expect the framework described here to prove useful for the experimental characterisation of work production in quantum heat engines~\cite{Lindenfels2019,Horne2020}, and, more generally, to help reveal and quantify the delicate fingerprints of genuinely quantum effects in non-equilibrium thermodynamic processes.

\acknowledgments
We acknowledge funding from a European Research Council Starting Grant ODYSSEY (Grant Agreement No.~758403). F.C.B.\ acknowledges funding from the European Union’s Horizon $2020$ research and innovation programme under the Marie Skłodowska-Curie Grant Agreement No.\ $801110$ and the Austrian Federal Ministry of Education, Science and Research (BMBWF). J. G. also acknowledges funding from a SFI Royal Society University Research Fellowship.

\appendix

\section{$\mathcal E_c$ is not a coherence monotone} 

In this Appendix we show that $\mathcal E_c$ is not a coherence monotone, i.e., there is some incoherent operation $\Lambda$ such that $\mathcal E_c(\Lambda(\hat \rho))\nleq \mathcal E_c(\hat \rho)$. We focus on the qubit example from the main text with  for which, choosing $\varepsilon_2=1$,
\begin{equation}
{\cal E}_c(\hat{ \rho}) = \frac{1}{2}\left( \sqrt{2 p(\hat{ \rho})-1}-\sqrt{2 p(\hat{ \rho})-1-C_{l_1}^2(\hat{ \rho})}\right)
\end{equation}

\begin{figure}
  \centering
  \includegraphics[width=0.8\columnwidth]{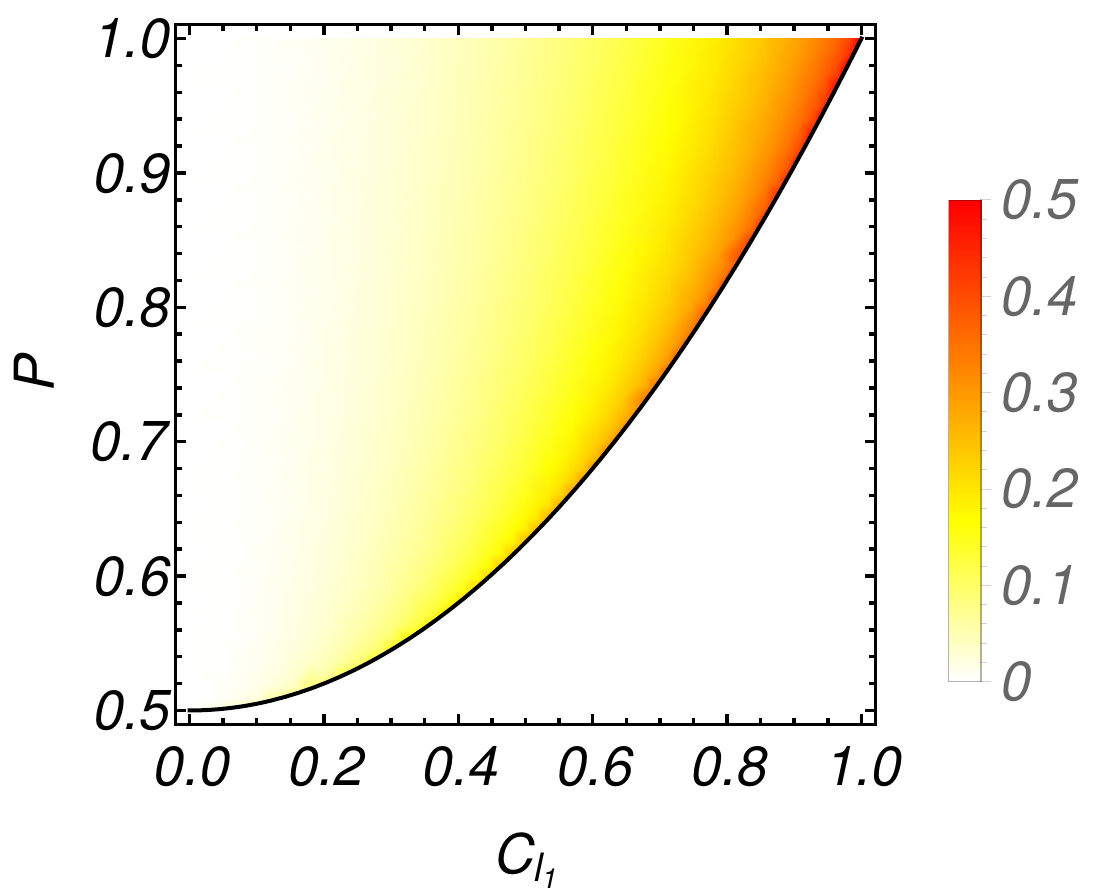}
  \caption{The density plot of the work $\mathcal{E}_c$ as a function of the purity $p(\hat \rho)$ and the coherence monotone $C_{l_1}$. The black line is $p(\hat \rho)=(1+C_{l_1}^2)/2$.
  }\label{fig:battery}
\end{figure}

An operation $\Lambda$ is an incoherent operation (IO) if it can be represented in terms of Kraus operators $K_i$ such that $K_i\hat \rho_i K_i^\dag$ is proportional to an incoherent state for all $i$ and incoherent inputs $\hat \rho_i$~\cite{cohrev}. For any such $\Lambda$ we have that $C_{l_1}(\Lambda(\hat{ \rho}))\leq C_{l_1}(\hat{ \rho})$. In contrast, as observed in the main text, if the purity $p(\Lambda(\hat{ \rho}))$ is smaller than $p_c\equiv\frac{1}{2}+\frac{1}{2}\left({\cal E}_c(\hat{ \rho})+\frac{C_{l_1}^2(\Lambda(\hat{ \rho}))}{4{\cal E}_c(\hat{ \rho})}\right)^2$ it turns out that $\mathcal E_c(\Lambda(\hat \rho))> \mathcal E_c(\hat \rho)$ (see Fig.~\ref{fig:battery}).

As an example of incoherent operations, we now consider the generalized amplitude damping map
\begin{equation}
\Omega(\hat \rho)=\sum E_j \hat \rho E_j^\dagger
\end{equation}

with Kraus operators

\begin{align*}
E_0 =& \sqrt{q} \begin{pmatrix} 1 & 0\\ 0 & \sqrt{1-\gamma} \end{pmatrix}, \\ E_1 =& \sqrt{q} \begin{pmatrix} 0 & \sqrt{\gamma} \\ 0 & 0 \end{pmatrix}, \\ E_2 =& \sqrt{1 - q} \begin{pmatrix} \sqrt{1-\gamma} & 0\\ 0 & 1 \end{pmatrix}, \\ E_3 =& \sqrt{1 - q} \begin{pmatrix} 0 & 0 \\ \sqrt{\gamma} & 0 \end{pmatrix}
\end{align*}\\

$\Omega$ maps an initial state $\hat \rho=\begin{pmatrix}  \rho_{11} &  \rho_{12}\\  \rho_{21} &  \rho_{22} \end{pmatrix}$ into
\begin{equation}\begin{split}
\Omega(\hat \rho)= (1-\gamma) \begin{pmatrix}  \rho_{11} & 0 \\ 0 & \rho_{22} \end{pmatrix} &+ \sqrt{1-\gamma} \begin{pmatrix} 0 &  \rho_{12} \\  \rho_{21} & 0\end{pmatrix}\\&+ \gamma \begin{pmatrix} q & 0 \\ 0 & 1 - q \end{pmatrix}.
\end{split}
\end{equation}

Here, we have separated the coherent and incoherent contribution from the state $\hat \rho$, as well as the state-independent contribution.

\begin{figure}
\centering
\includegraphics[width=0.8\columnwidth]{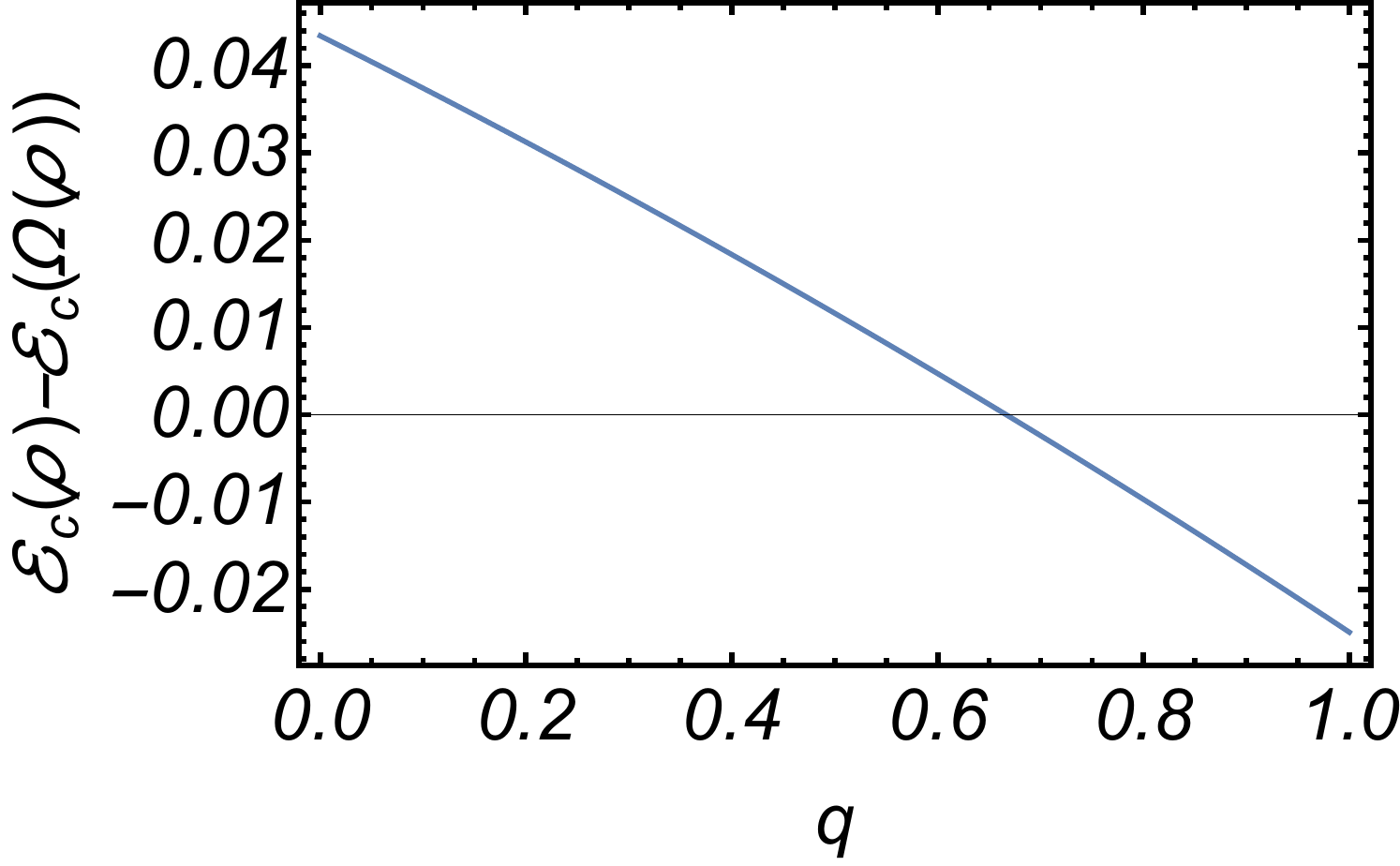}
\caption{The plot shows the difference $\mathcal E_c(\hat \rho)-\mathcal E_c(\Omega(\hat \rho))$ as a function of the parameter $q$. }\label{fig}
\end{figure}

In Fig.~\ref{fig} we study $\mathcal E_c(\hat \rho)-\mathcal E_c(\Omega(\hat \rho))$ as a function of the parameter $q$, for $\gamma=1/10$, $ \rho_{11}=1/3$ and $ \rho_{12}=\sqrt{ \rho_{11} \rho_{22}}$. Notice that when  the purity $p(\Omega(\hat{ \rho}))$ is smaller than $p_c$, we have that $\mathcal E_c(\Omega(\hat \rho))>\mathcal E_c(\hat \rho)$. Therefore, $\mathcal E_c$ is not a coherence monotone under incoherent operations.

\vfill


\begin{thebibliography}{99}
\bibitem{degeneracy}
If the Hamiltonian has degenerate levels, the passive state is defined up to unitaries acting in each degenerate subspace. Through such a unitary, it is possible to change the coherence without altering the ergotropy, which, thus, is independent of the coherence between degenerate states. To avoid counting this kind of coherence in $C(\hat{\rho})$, we can always choose the energy eigenbasis such that $\hat{\rho}$ is diagonal in every  degenerate subspace.
\bibitem{fixperm}
Just to fix the notation for the permutation, let $d=4$, and take $\pi$ such that
$\pi(\{1,2,3,4\}) = \{3,2,4,1\}$. In this case, we write $\pi_1= 3, \pi_2=2, \pi_3=4,\pi_4=1$.
The inverse permutation, $\pi^{-1}$, is the one such that $\pi^{-1}(\pi( \{1,2,3,4\}) ) = \{1,2,3,4\}$.
Thus, in this case, $\pi^{-1}(\{1,2,3,4\}) = \{4,2,1,3\}$, and $\pi^{-1}_1=4, \pi^{-1}_2=2, \pi^{-1}_3=1, \pi^{-1}_4=3$.
It follows that, if $\pi_k= k'$, then, $k=\pi^{-1}_{k'}$.
\bibitem{nota}
We note that, when the state $\hat \rho$ does not commute with $H$ it evolves in the time. Thus, the application of the ergotropic unitary cyle $\hat E$ requires a very precise timing. On the other hand, this is not the case for the incoherent operation $\hat V_{\pi}$.
\end{thebibliography}
\end{document}